\pdfoutput=1
\documentclass[aps,prd,amsmath,floats,floatfix, twocolumn,
superscriptaddress,nofootinbib,showpacs,longbibliography]{revtex4-1}
\usepackage[T1]{fontenc}
\usepackage[utf8]{inputenc}
\usepackage{txfonts}
\usepackage{verbatim}
\usepackage[dvipsnames, usenames]{xcolor}
\definecolor{linkcolor}{rgb}{0.0,0.3,0.5}
\usepackage[hypertexnames=false, unicode, colorlinks=true, linkcolor=linkcolor,
citecolor=linkcolor, filecolor=linkcolor,urlcolor=linkcolor,
pdfusetitle]{hyperref}
\usepackage[all]{hypcap}
\usepackage{graphicx}
\usepackage{xspace}
\usepackage{amssymb}
\usepackage{microtype}
\usepackage{subfigure}

\usepackage{url}

\usepackage[english]{babel}
\usepackage{blindtext}

\usepackage[normalem]{ulem} 
\usepackage{bm} 
\usepackage{mathrsfs}
\usepackage{xspace} 
\usepackage{fontawesome}
\usepackage[normalem]{ulem}

\DeclareMathAlphabet{\mathpzc}{OT1}{pzc}{m}{it}

\usepackage[nomessages]{fp}

\newcommand{\sk}[1]{}

\newcommand{\gwModel}{\textcolor{linkcolor}{\texttt{gwModel}}}
\newcommand{\gwModelX}{\textcolor{linkcolor}{\texttt{gwModel\_kick\_q200}}}
\newcommand{\gwModelXP}{\textcolor{linkcolor}{\texttt{gwModel\_prec\_flow}}}
\newcommand{\HLZ}{\textcolor{linkcolor}{\texttt{HLZ}}}
\newcommand{\NRSur}{\textcolor{linkcolor}{\texttt{NRSur}}}

\begin{document}
\title{Progenitor of the recoiling super-massive black hole RBH-1 identified using HST/JWST imaging}

\author{Tousif Islam}
\email{tousifislam@ucsb.edu}
\affiliation{Kavli Institute for Theoretical Physics,University of California Santa Barbara, Kohn Hall, Lagoon Rd, Santa Barbara, CA 93106}

\author{Tejaswi Venumadhav}
\affiliation{\mbox{Department of Physics, University of California at Santa Barbara, Santa Barbara, CA 93106, USA}}
\affiliation{\mbox{International Centre for Theoretical Sciences, Tata Institute of Fundamental Research, Bangalore 560089, India}}

\author{Digvijay Wadekar}
\affiliation{Department of Physics and Astronomy, Johns Hopkins University, 3400 N. Charles Street, Baltimore, Maryland, 21218, USA}
\affiliation{\mbox{Weinberg Institute, University of Texas at Austin, Austin, TX 78712, USA}}

\hypersetup{pdfauthor={Islam et al.}}

\date{\today}

\begin{abstract}
Using a combination of \textit{Hubble Space Telescope} and \textit{James Webb Space Telescope} imaging, a runaway supermassive black hole (RBH-1) was recently identified with an inferred velocity of $954^{+110}_{-126},\mathrm{km,s^{-1}}$, likely ejected from a compact star-forming galaxy (denoted GX) at $z \approx 0.96$. Assuming the runaway black hole originated from a gravitational-wave-driven merger of two SMBHs, we combine its measured recoil velocity with gravitational-wave recoil predictions from numerical relativity and black-hole perturbation theory to constrain the mass ratio and spin configuration of the progenitor binary that overcame the final-parsec problem and merged $\sim 70$ Myr ago. We find that the progenitor binary must have been precessing, with a mass ratio $m_1/m_2 \lesssim 6$, and that the more massive SMBH likely possessed a high dimensionless spin magnitude ($\sim 0.75$) in order to generate a recoil of this magnitude. Such SMBH mergers could represent an interesting source population for the upcoming LISA mission, with characteristic signal-to-noise ratios of order $\gtrsim 10^3$. Furthermore, the inferred progenitor SMBH properties suggest that GX likely originated from a major, gas-rich (“wet”) merger between two galaxies of comparable mass, with a mass ratio $\lesssim 4$.
\end{abstract}

\maketitle

\noindent \textbf{\textit{\textit{Introduction: }}} Using deep \textit{Hubble Space Telescope} (HST) imaging, Ref.~\cite{vanDokkum:2023wed} identified a remarkably thin, $62$~kpc--long linear feature at redshift $z=0.96$ that points directly to the center of a compact galaxy (GX), indicating a physical association~\cite{vandokkum2023directconnectionwakehost}. The structure has an almost constant $\sim 1$~kpc width, blue rest-frame colors, and numerous clumps of recent star formation. At its leading end lies an unresolved point-like source with no detectable stellar continuum, moving at supersonic speed relative to GX~\cite{vanDokkum:2023wed}. The combination of a thin star-forming wake and a compact, featureless apex is unlike any known tidal tail, radio jet, or galactic disk. Ref.~\cite{vanDokkum:2023wed} interpreted the system as a supermassive black hole (SMBH) ejected from GX and traversing the circumgalactic medium (CGM), where it would drive a bow shock and trigger in-situ star formation along its path.
Ref.~\cite{vanDokkum:2025bah} subsequently used JWST Near-Infrared Spectrograph (NIRSpec) Integral Field Unit (IFU) spectroscopy, together with supporting HST ultraviolet-visible (UVIS) imaging, to map the gas kinematics and excitation across the structure. The apex displays a sharp velocity discontinuity of $\sim 600~\mathrm{km , s^{-1}}$, characteristic of a bow shock, while the downstream gas shows a smooth velocity gradient and line ratios consistent with shock-excited, cooling material. The observed gradient and projected post-shock flow speed of $\sim 300~\mathrm{km , s^{-1}}$ are well reproduced by a simple shock-compression model of a supersonic object with velocity $v_\bullet = 954^{+110}_{-126}~\mathrm{km,s^{-1}}$, corresponding to an ejection event $\sim 70$ Myr ago~\cite{vanDokkum:2025bah}. The interpretation is further supported by the morphology near the wake tip and by diagnostic ratios such as [O III]/H$\alpha$, [N II]/H$\alpha$, [S II]/H$\alpha$, and [S III]/[S II], which indicate fast radiative shocks consistent with the inferred motion and shock geometry.
Alternative explanations such as a bulgeless edge-on galaxy~\cite{montes2024deephstimagingfavors} or a tidal feature~\cite{vanDokkum:2025bah} cannot reproduce the absence of stellar continuum at the tip, the extreme velocity, or the required mechanical energy injection, effectively ruling out known stellar or galactic perturbers. Because the bow continuously thermalizes kinetic energy as the supersonic object moves through the CGM, the observed velocity, wake size ($\sim 1.2~\mathrm{kpc}$), and typical CGM density ($10^{-3}$ $\rm cm^{-3}$) require a mass of $M_\bullet \gtrsim 10^7,M_\odot$, incompatible with stellar remnants or clusters and naturally pointing to an SMBH (RBH-1 hereafter).

\noindent \textbf{\textit{\textit{Gravitational-wave recoil interpretation: }}} RBH-1 can be explained through two main mechanisms. One possibility is a gravitational-wave (GW) recoil~\cite{Bonnor1961,BreuerHehl1962,1973ApJ183657B} produced when two SMBHs, originally hosted in the progenitor galaxies of GX, merged after overcoming the final-parsec barrier. Most galaxies are expected to undergo at least one major merger during their lifetime~\cite{2021MNRAS5013215O}. Alternatively, a three-body interaction with a second SMBH could eject one object at high velocity. However, the low stellar velocity dispersion of GX ($\sim 60~\mathrm{km,s^{-1}}$) makes such encounters unlikely, and no evidence has been found for the second runaway SMBH expected in this scenario~\cite{vanDokkum:2025bah}. Numerical-relativity (NR) simulations, by contrast, show that GW recoil velocities can exceed $\sim 5000~\mathrm{km,s^{-1}}$~\cite{Campanelli:2007cga,Bruegmann:2007bri,Campanelli:2007ew,Choi:2007eu,Dain:2008ck,Gonzalez:2006md,Gonzalez:2007hi,Healy:2008js,Healy:2022jbh}, easily exceeding the measured motion of the compact source. The parameters inferred for GW merger events GW200129 and GW191109 in the LIGO--Virgo--KAGRA data also indicate recoil velocities of $\sim 1500~\mathrm{km,s^{-1}}$ and $\sim 500~\mathrm{km,s^{-1}}$, respectively, showing that large recoils occur in nature, albeit in a different mass regime~\cite{Varma:2022pld,Islam:2023zzj}. Thus, a GW recoil from an SMBH merger provides the most natural explanation for RBH-1.

Several runaway or recoiling SMBH candidates have previously been proposed based on astrometric or spectroscopic searches for active galactic nuclei (AGN) displaced from their galactic centers~\cite{Komossa:2008qd,Chiaberge:2025ouh,Shields:2008kn,2014MNRAS4451558D,2009ApJ707936S,2013MNRAS4281341B,2025ApJ99238B,2017A&A600A57C}. Many were later ruled out or reinterpreted. In contrast, RBH-1 is the most compelling recoiling-SMBH candidate to date because of its well-measured 62~kpc shock-induced star-forming wake, IFU-measured gas kinematics, and well-resolved bow-shock geometry~\cite{vanDokkum:2023wed,vanDokkum:2025bah}. Assuming the recoiling-SMBH interpretation, we show that the runaway velocity already constrains several properties of the progenitor SMBH binary, including its mass ratio and spin configuration, and discuss the implications for SMBH mergers and the progenitor system of GX.

\begin{figure}
    \centering
    \includegraphics[width=\columnwidth]{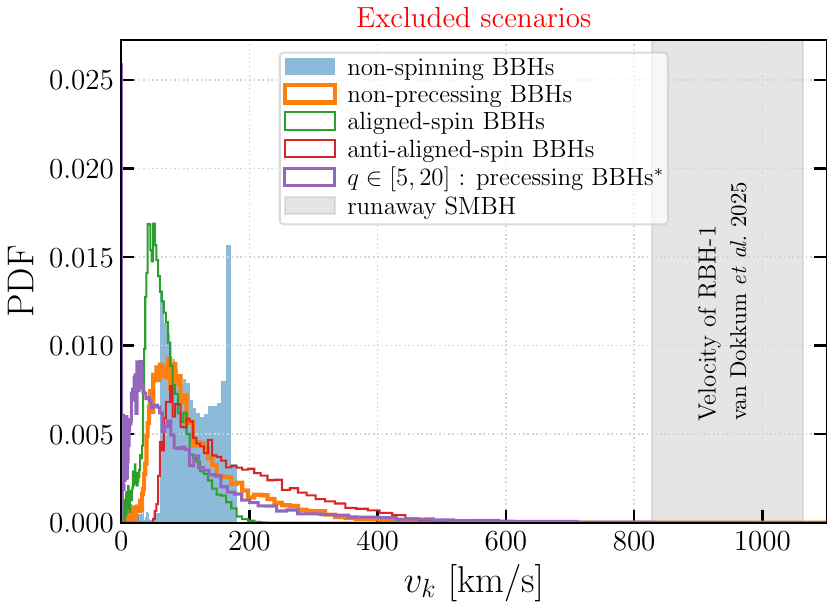}
    \caption{We show merger scenarios \emph{incapable} of producing the RBH-1 runaway velocity of $954^{+110}_{-126}\,\mathrm{km\,s^{-1}}$ (vertical shaded gray region). These include non-spinning SMBH mergers, non-precessing SMBH mergers, and precessing SMBH mergers with mass ratios $q \in [5,20]$. Only near-equal-mass precessing systems, in which the SMBH spins are misaligned with the orbital angular momentum, can produce such large recoils. We use \gwModel{} for this figure.}
    \label{fig:progenitor_types}
\end{figure}

\begin{figure}
    \centering
    \includegraphics[width=\columnwidth]{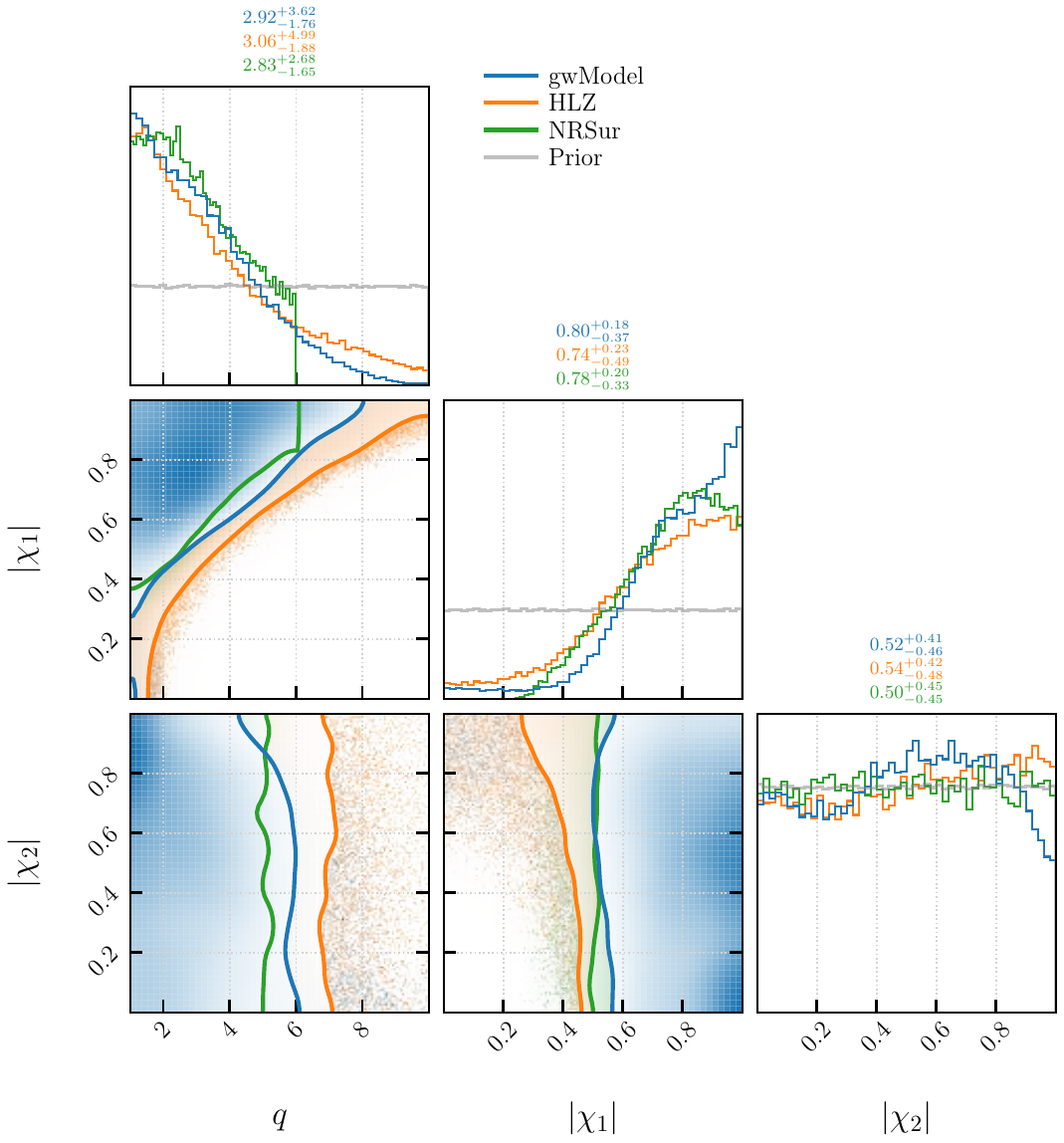}
    \caption{We show the progenitor SMBH mass ratio $q (=m_1/m_2)$ and dimensionless spin magnitudes $|\chi_{1,2}|$ consistent with the inferred RBH-1 runaway velocity of $954^{+110}_{-126}\,\mathrm{km\,s^{-1}}$, assuming a precessing progenitor binary. We use three recoil-kick models calibrated against NR and/or BHPT simulations: \gwModel{} (blue), \HLZ{} (orange), and \NRSur{} (green). The priors are shown in gray. There is a clear preference for more symmetric binaries with a rapidly spinning primary SMBH.}
    \label{fig:progenitor}
\end{figure}

\noindent \textbf{\textit{Models for gravitational recoil: }} Our default model, \gwModel{}~\cite{Islam:2025drw}, is a recoil-kick prescription calibrated to numerical-relativity (NR) and black-hole perturbation theory (BHPT) simulations for binary mass ratios \(q \equiv m_1/m_2 \leq 200\) and dimensionless spin magnitudes \(|\chi_{1,2}| \leq 1\). The model is analytical and deterministic in the aligned-spin limit (\gwModelX{}), while for precessing binaries it uses a probabilistic normalizing-flow framework (\gwModelXP{}) that marginalizes over spin angles and returns a kick-velocity distribution rather than a point estimate. We use the implementation provided in the \textcolor{linkcolor}{\texttt{gwModels}} package\footnote{\href{https://github.com/tousifislam/gwModels}{https://github.com/tousifislam/gwModels}}.
To assess model systematics, we also use two recoil prescriptions calibrated with different methods and datasets. The first is \HLZ{}~\cite{Lousto:2008dn,Lousto:2010xk,Lousto:2012gt,Lousto:2012su,Gonzalez:2007hi}, a semi-analytical fit constructed from a smaller NR dataset. The second is \textcolor{linkcolor}{\texttt{NRSur7dq4Remnant}}~\cite{Varma:2019csw} (hereafter \textcolor{linkcolor}{\texttt{NRSur}}), a Gaussian-process surrogate trained on \(\sim 1500\) NR simulations within \(q \leq 4\) and \(|\chi_{1,2}| \leq 0.8\), but which can be reasonably extrapolated to \(q \leq 6\) and \(|\chi_{1,2}| \leq 1\). For the spin angles, we use a reference frame close to merger (details are provided in the Supplemental Material).

\noindent \textbf{\textit{Constraining progenitor SMBH merger types: }} First, we determine which merger configurations are incapable of producing the runaway velocity \(v_{\rm BH} = 954^{+110}_{-126}\,\mathrm{km\,s^{-1}}\). Using \gwModel{}, we compute recoil velocities for non-spinning BBHs with mass ratios sampled from \(q \in [1,20]\). We find that the kick velocity never exceeds \(\sim 180~\mathrm{km\,s^{-1}}\) (Fig.~\ref{fig:progenitor_types}). For \(q \gtrsim 20\), the recoil decreases monotonically with mass ratio, so extending to larger \(q\) cannot produce higher kicks.
Next, we consider BBHs without precession. For aligned-spin binaries within the same mass-ratio range, the maximum recoil is \(\sim 200~\mathrm{km\,s^{-1}}\). When the BH spins are anti-aligned with the orbital angular momentum, the maximum recoil increases to \(\sim 300~\mathrm{km\,s^{-1}}\), still far below the inferred runaway velocity (Fig.~\ref{fig:progenitor_types}). These results strongly disfavor progenitor SMBH mergers involving either non-spinning or non-precessing binaries.

\begin{figure}
    \centering
    \includegraphics[width=\columnwidth]{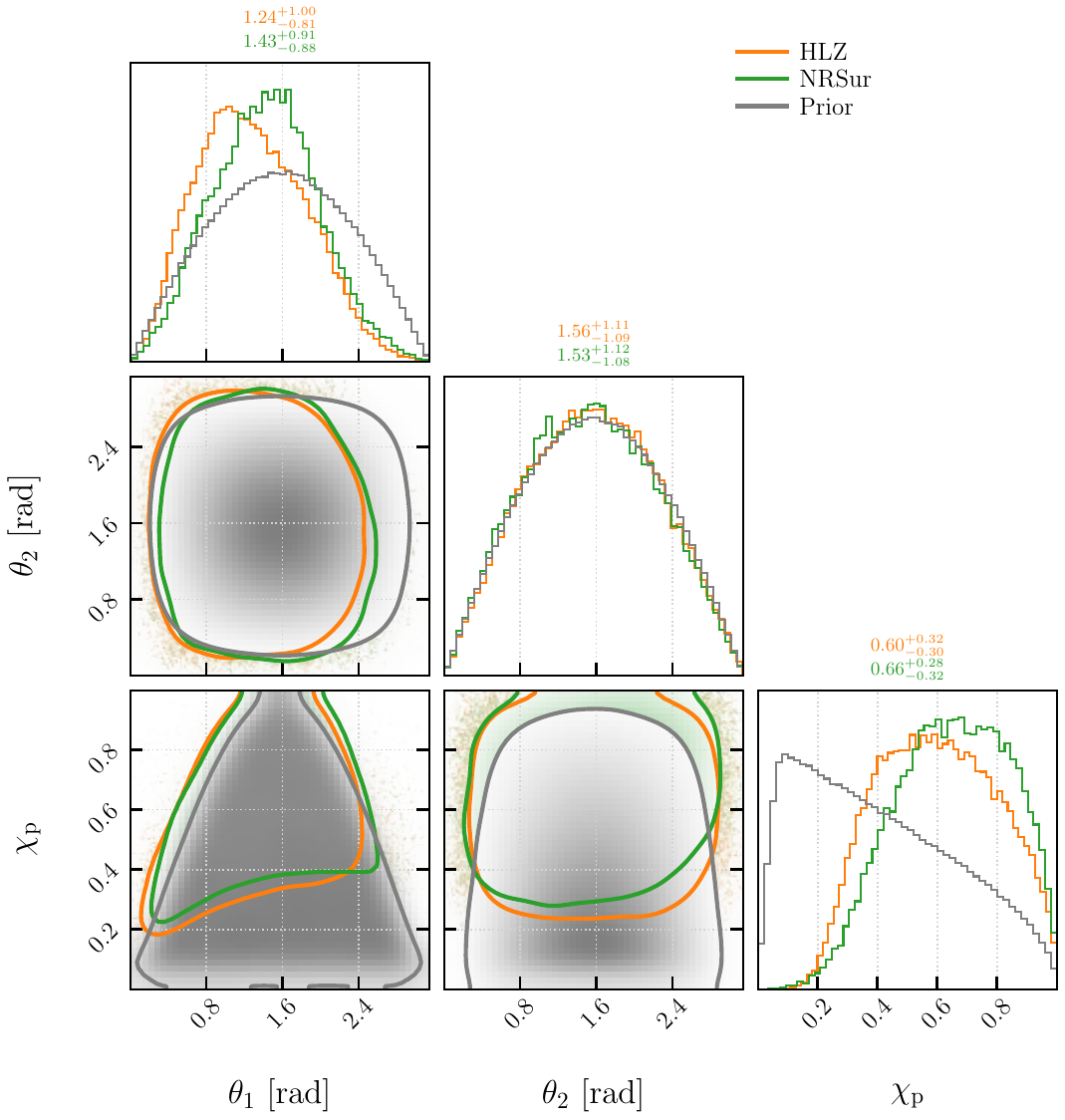}
    \caption{Similar to Fig.~\ref{fig:progenitor}, but showing the inferred spin angles between the progenitor SMBH spins and the orbital angular momentum: $\theta_1\ (=\cos^{-1}[\hat{L}\cdot\hat{S}_1])$, $\theta_2$, and the effective precession parameter $\chi_{\rm p}$ (Eq.~\ref{eq:chi_p}), obtained using the recoil-kick models \HLZ{} and \NRSur{}. We do not show \gwModel{} because it marginalizes over spin angles internally. The priors are shown in gray. There is a strong preference for spin precession and a weak constraint on the primary-spin angle.}
    \label{fig:progenitor_precession}
\end{figure}

This analysis therefore implies that the two SMBHs involved in the merger had substantial in-plane spin components, i.e., the binary must have been precessing. We then ask whether precessing SMBHs of \emph{any} mass ratio could generate such a large recoil. To this end, we sample mass ratios from \(q \in [5,20]\), draw dimensionless spin magnitudes uniformly from \(|\chi_{1,2}| \leq 1\), and assume isotropic spin-orientation angles \((\theta_{1},\theta_{2},\phi_{1},\phi_{2})\). The angles \(\theta_{1}\) and \(\theta_{2}\) denote the tilt angles between the individual SMBH spin directions \(\hat{S}_{1,2}\) and the orbital angular momentum \(\hat{L}\), while \(\phi_{1}\) and \(\phi_{2}\) specify the corresponding in-plane spin angles.
We find that the resulting kick-velocity distribution remains strongly peaked around \(\sim 200~\mathrm{km\,s^{-1}}\), making it essentially impossible to generate a recoil near \(v_{\rm BH} = 954^{+110}_{-126}\,\mathrm{km\,s^{-1}}\) in this mass-ratio regime (Fig.~\ref{fig:progenitor_types}). This implies that the progenitor SMBH binary capable of producing such a runaway remnant must have been a highly precessing system with mass ratio \(q \in [1,5]\). Although we present results here using \gwModel{}, \HLZ{} and \NRSur{} yield the same qualitative conclusions.

\noindent \textbf{\textit{Constraining the mass ratio and spins of the progenitor SMBHs:}} 
To constrain the mass ratio $q$, the dimensionless spin magnitudes $\chi_{1,2}$, and the spin-orientation angles $\{\theta_1,\theta_2,\phi_1,\phi_2\}$ of the progenitor SMBHs, we use Monte Carlo forward modeling with rejection sampling. Because the kick velocity forward model is extremely cheap and the parameter space is low-dimensional, rejection sampling gives independent accepted samples without burn-in or autocorrelation concerns.
We draw $500{,}000$ binary configurations, sufficient for convergence of the inferred distributions, uniformly in the ranges $q \in [1,10]$ and $|\chi_{1,2}| \in [0,1]$. For the \NRSur{} model, we restrict to $q \in [1,6]$ because its extrapolation beyond this range is unreliable~\cite{Islam:2025drw,Varma:2019csw}. Spin-orientation angles are drawn isotropically, i.e., $\cos\theta_{1,2}$ uniformly in $[-1,1]$ and $\phi_{1,2}$ uniformly in $[0,2\pi]$.
For each sampled binary, we compute three remnant quantities: the ratio of final to initial mass $m_f/m_i$, the final dimensionless spin magnitude $\chi_f$, and the recoil (kick) velocity $v_f$. These distributions constitute our \emph{prior}. We then select only those configurations whose kick velocities fall within the observed runaway-SMBH velocity range, $v_{\rm BH} = 954^{+110}_{-126}\,\mathrm{km\,s^{-1}}$. 
The surviving configurations define the accepted distribution; we refer to this conditional distribution as the \emph{posterior} for brevity.
We discuss the robustness of this approach in the Supplemental Material. We find that approximately $7\%$ of the prior yields a recoil velocity of at least $828~\mathrm{km\,s^{-1}}$ (the lower limit for RBH-1) when \gwModel{} is used. This fraction increases to $\sim 11.4\%$ and $\sim 15\%$ when using \HLZ{} and \NRSur{}, respectively.

The inferred priors and posteriors for the most important binary parameters ($q$, $|\chi_1|$, and $|\chi_2|$) are shown in Fig.~\ref{fig:progenitor}. We find that the runaway velocity strongly constrains the mass ratio to be  
$q = 2.91^{+3.61}_{-1.75}$ for \gwModel{},  
$q = 3.06^{+4.96}_{-1.88}$ for \HLZ{},  
and $q = 2.86^{+2.63}_{-1.67}$ for \NRSur{}.  
Furthermore, the recoil constraint requires the \emph{larger} progenitor SMBH to have a high dimensionless spin magnitude, with  
$\chi_{1} = 0.80^{+0.18}_{-0.38}$ for \gwModel{},  
$\chi_{1} = 0.75^{+0.23}_{-0.49}$ for \HLZ{},  
and $\chi_{1} = 0.78^{+0.20}_{-0.34}$ for \NRSur{}.  
In contrast, the dimensionless spin magnitude of the smaller progenitor SMBH remains essentially unconstrained.

We also find that most spin-precession angles are essentially unconstrained, with posterior distributions nearly identical to their priors. The only exception is $\theta_{1}$ (Fig.~\ref{fig:progenitor_precession}), whose posterior shows a modest but noticeable deviation from the prior and favors values corresponding to maximally precessing configurations. 
Furthermore, we compute the effective spin-precession parameter $\chi_{\rm p}$, defined as~\cite{Schmidt:2014iyl}
\begin{equation}
 \chi_{\rm p} = \max\!\left(\chi_{1}\sin\theta_{1},\,
\frac{4+3q}{q(4q+3)}\,\chi_{2}\sin\theta_{2}\right),
\label{eq:chi_p}
\end{equation}
which captures the dominant in-plane spin contribution driving precession. Binaries with no spin precession have $\chi_{\rm p}=0$, whereas maximally precessing binaries have $\chi_{\rm p}=1$. We find that the constraints on $q$ and $\theta_{1}$ produce a posterior on $\chi_{\rm p}$ that differs significantly from the prior, indicating that the observed runaway velocity carries nontrivial information about the precessional dynamics of the progenitor SMBH binary.
Because \gwModel{} marginalizes over spin angles internally, this part of the analysis is carried out using \HLZ{} and \NRSur{}.

Ref.~\cite{vanDokkum:2023wed} does not constrain the spin of RBH-1. Under the gravitational-recoil interpretation, however, the correlation between remnant spin and recoil velocity allows us to infer the dimensionless spin magnitude of RBH-1. From our analysis in Fig.~\ref{fig:progenitor}, we infer
$\chi_{\rm SMBH} = 0.70^{+0.20}_{-0.35}$ using \gwModel{},
$\chi_{\rm SMBH} = 0.62^{+0.23}_{-0.30}$ using \HLZ{}, and
$\chi_{\rm SMBH} = 0.72^{+0.16}_{-0.19}$ using \NRSur{} (using the remnant-spin fit of Ref.~\cite{2009ApJ...704L..40B}).
We also find that the observed runaway velocity requires the ratio of final to initial binary mass to cluster tightly around $\sim 0.97$ (using the remnant-mass fit of Ref.~\cite{Barausse:2012qz}). The inferred progenitor properties are broadly similar to those of the runaway-SMBH candidate quasar 3C~186 at $z=1.068$, which exhibits a recoil velocity of $1310^{+21}_{-21},\mathrm{km,s^{-1}}$~\cite{2017A&A600A57C,Chiaberge:2025ouh}. Further discussion is provided in the Supplemental Material.

\noindent \textbf{\textit{Astrophysical implications: }}
Combining constraints from the three recoil--kick prescriptions, we find a characteristic mass ratio \(q \simeq 3\), with a robust lower bound \(q \gtrsim 1.2\) and an upper bound \(q \lesssim 6\text{--}7\) for the progenitor SMBHs. 
We use an empirical scaling relation between SMBH mass ($M_{\rm SMBH}$) and host-galaxy stellar or bulge mass ($M_{\rm bulge}$)~\cite{Kormendy:2013dxa,2019ApJ887245S}:
\[
\log\!\left(\frac{M_{\rm SMBH}}{M_\odot}\right)
= (1.24 \pm 0.08)\,
\log\!\left(\frac{M_{\rm bulge}}{10^{11} M_\odot}\right)
+ (8.80 \pm 0.09)\,.
\]
This suggests that the galaxies that merged to form GX also had a comparable mass ratio, \(1 \lesssim q_{\rm gal} \lesssim 4\), placing the event in the major- or near-major-merger regime~\cite{2025arXiv250609136K}.

\begin{figure}
    \centering
    \includegraphics[width=\columnwidth]{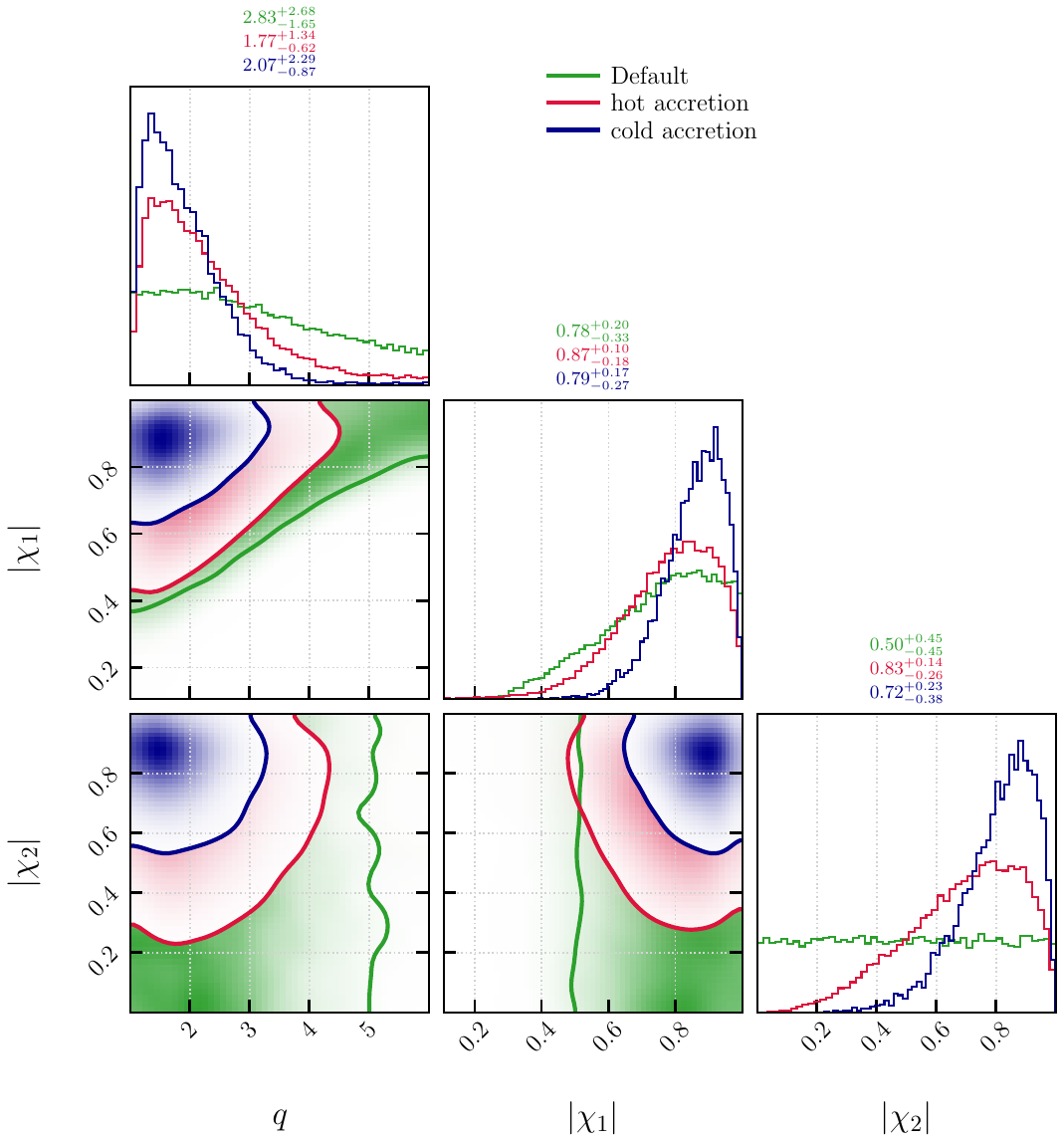}
    \caption{Same as Fig.~\ref{fig:progenitor} but using astrophysically-motivated priors from two different accretion scenarios in gaseous environments~\cite{Lousto:2017uav,Lousto:2012su,Zlochower:2015wga}: hot accretion (crimson) and cold accretion (dark blue), while the default model is shown again for comparison. We use the \NRSur{} recoil kick model for this plot.}
    \label{fig:accretion_models}
\end{figure}

The inferred precession of the progenitor SMBH binary requires spin--orbit misalignment, indicating a dynamically complex, gas-rich environment rather than a gas-poor or purely collisionless merger. Gas-rich environments can also help overcome the final-parsec problem through stellar and gaseous torques~\cite{Nixon:2013qfa,2006MNRAS373L90K,2012MNRAS.425.1121H,2009MNRAS.398.1392L}.
Furthermore, the inferred spin of at least one SMBH, \(\chi \sim 0.8\), is more naturally associated with prolonged gas accretion~\cite{Thorne:1974ve,Berti:2008af} than with repeated minor mergers or purely chaotic assembly histories~\cite{Hughes:2002ei,Berti:2008af,Sesana:2007sh,2025ApJ...991...58K,Kritos:2024kpn}.
This combination of high spin and strong precession is, at face value, in tension
with the standard picture of coherent gas accretion in a gas-rich environment,
which would align both spins with the orbital angular momentum via the
Bardeen-Petterson effect~\cite{1975ApJ...195L..65B} and cap recoils below
the observed value. The tension is alleviated if the spin magnitudes were
largely set in each progenitor galaxy \emph{before} the galaxy merger, via coherent accretion in each pre-merger nucleus,
while the spin orientations at merger are determined by the much shorter
circumbinary phase that follows. Incomplete alignment during this
circumbinary phase can arise from warped or torn circumbinary disks above a
critical obliquity~\cite{Nealon:2021akj,deSimone:2025cqg}, hot or thick
accretion flows with weak warp propagation, or alignment timescales
comparable to or longer than the gas-driven inspiral
timescale~\cite{Bogdanovic:2007hp,Dotti:2012qw}. The hot accretion scenario
shown in Fig.~\ref{fig:accretion_models} populates exactly this
intermediate regime and is consistent with our posteriors.
Taken together, these considerations favor a major, gas-rich merger origin for GX.

One possibility is that the progenitor binary resided in a nuclear star cluster, but sustained gas accretion capable of producing the large inferred SMBH spins is less likely in such systems. Formation within a gaseous disk is more naturally associated with large SMBH spins and gas-driven inspiral~\cite{Lousto:2017uav,Lousto:2012su,Zlochower:2015wga,Schnittman:2015eaa,Gerosa:2015xya,ColemanMiller:2013jrk,Blecha:2010dq}. Such environments could also produce pre-merger active galactic nucleus (AGN) activity and fossil signatures in the host galaxy~\cite{deSimone:2025cqg,2024MNRAS.52711233K}. In particular, precessing AGN jets associated with the progenitor SMBH binary could generate disturbed radio morphology or curved radio lobes prior to merger~\cite{1985ApJ...294L..85O,Murgia:2000tn}. Such features are expected to fade on the $\sim 70$ Myr timescale associated with RBH-1, consistent with current Very Large Array (VLA) and \textit{Chandra} non-detections. The host galaxy also shows disturbed optical morphology, potentially consistent with either a recent galaxy merger or fossil AGN activity~\cite{vanDokkum:2023wed}. Because AGN outflow signatures may persist longer than jets, future JWST/NIRSpec IFU or VLT-MUSE observations could search for extended gas and stellar bulk motions. The recoil velocity alone, however, cannot constrain the jet direction relative to the recoil direction or the total angular momentum of the progenitor SMBH binary (Supplemental Material).

If there is a population of similar precessing SMBH mergers in gaseous environments, they could be interesting multi-messenger sources for the planned LISA mission~\cite{LISA:2022yao}. For reference, the merger producing RBH-1 would have yielded a characteristic LISA signal-to-noise ratio (SNR) of order $\sim 10^3$ using our maximum-likelihood parameters and $M_\mathrm{SMBH}\sim 2\times10^7,M_\odot$ inferred from the host-galaxy scaling relation~\cite{vanDokkum:2025bah}. More details are in the Supplemental Material.

Our GW-recoil inferences complement constraints on RBH-1 from scaling relations and gas kinematics near the runaway SMBH~\cite{vanDokkum:2023wed,vanDokkum:2025bah}. Recoil velocities are independent of the total binary mass in general relativity, but if the recoiling SMBH mass is indeed $\mathcal{O}(10^7,M_\odot)$, drag from the circumgalactic medium would imply that the BH has slowed down since merger. In that case, the progenitor binary was even closer to equal mass and/or more strongly precessing at merger than inferred from the currently observed recoil velocity (Supplemental Material).

\noindent \textbf{\textit{Physically motivated priors:}}
Motivated by the results above, we repeat the analysis with physically motivated priors following Refs.~\cite{Lousto:2017uav,Lousto:2012su,Zlochower:2015wga,Schnittman:2015eaa,Gerosa:2015xya,ColemanMiller:2013jrk,Blecha:2010dq,Yu:2011vp,Hopkins:2009yy,Stewart:2008ep}. Based on cosmological simulations, the mass ratio is sampled from the distribution \(p(q)\propto (1/q)^{-0.3}(1-1/q)\)~\cite{Lousto:2017uav,Lousto:2012su,Zlochower:2015wga,Yu:2011vp,Hopkins:2009yy,Stewart:2008ep}. 
For spin magnitudes \(\chi_{1,2}\) and tilt angles \(\theta_{1,2}\), we consider two accretion-driven models~\cite{Lousto:2017uav,Lousto:2012su,Zlochower:2015wga,Schnittman:2015eaa,Gerosa:2015xya,ColemanMiller:2013jrk,Blecha:2010dq}. In gas-rich environments, hot accretion produces partial spin alignment and non-zero spin magnitudes, while cold accretion drives \(\chi_{1,2}\) toward \(1\) and yields tightly aligned spins. In both cases, the distributions of \(\chi_{1,2}\) and \(\theta_{1,2}\) are modeled as Beta distributions with different exponents~\cite{Lousto:2017uav,Lousto:2012su,Zlochower:2015wga}.
We find that these astrophysical priors favor progenitor SMBH binaries that are more equal mass and more rapidly spinning (Fig.~\ref{fig:accretion_models}), although the qualitative conclusions remain unchanged. While we use \NRSur{} for this analysis, the results remain similar when using \HLZ{}.

\noindent \textbf{\textit{Acknowledgment: }}We thank Jessica Lu, Crystal Martin, Mathieu Renzo, Marc Favata, Javier Roulet, Pieter van Dokkum, and the anonymous referees for insightful discussions, careful reading of the manuscript, and constructive suggestions.
T.I. is supported in part by the National Science Foundation under Grant No. NSF PHY-2309135 and the Gordon and Betty Moore Foundation Grant No. GBMF7392. 
TV acknowledges support from NSF grants 2012086 and 2309360, the Alfred P. Sloan Foundation through grant number FG-2023-20470, the BSF through award number 2022136, and the Hellman Family Faculty Fellowship. 
D.W. is supported by NSF Grants No.~AST-2307146, No.~PHY-2513337, No.~PHY-090003, and No.~PHY-20043, by NASA Grant No.~21-ATP21-0010, by John Templeton Foundation Grant No.~62840, by the Simons Foundation [MPS-SIP-00001698, E.B.], by the Simons Foundation International [SFI-MPS-BH-00012593-02], and by Italian Ministry of Foreign Affairs and International Cooperation Grant No.~PGR01167.

\bibliography{kick_references}

\begin{figure}
    \centering
    \includegraphics[width=\columnwidth]{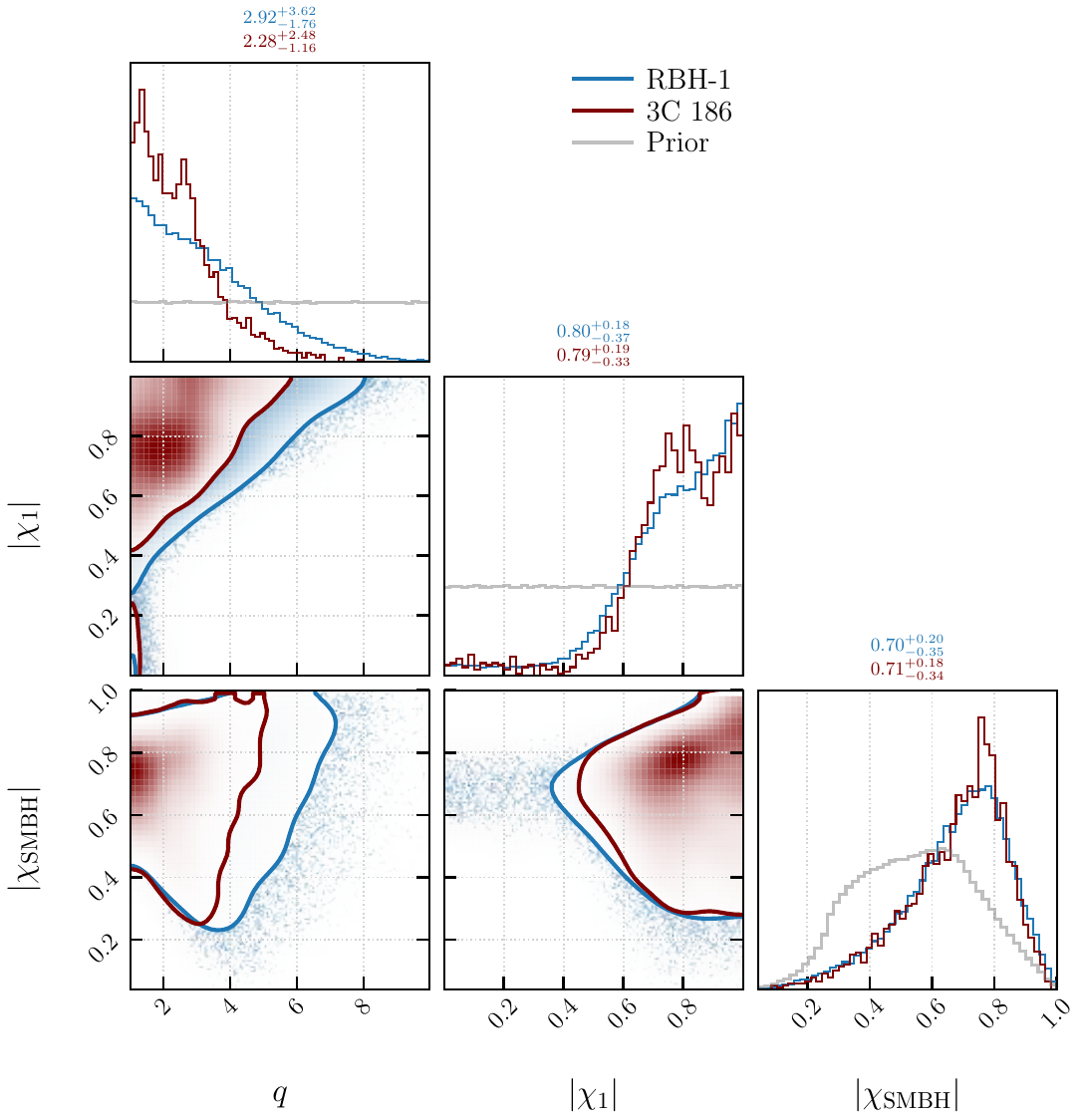}
    \caption{We show the progenitor SMBH mass ratio $q$, the dimensionless spin magnitude of the larger progenitor SMBH $|\chi_{1}|$, and the resulting remnant SMBH dimensionless spin magnitude $\chi_{\rm SMBH}$ consistent with the inferred runaway velocities of $954^{+110}_{-126},\mathrm{km,s^{-1}}$ and $1310^{+21}_{-21},\mathrm{km,s^{-1}}$ for RBH-1 (blue) and the quasar 3C~186 (brown), respectively, under the assumption that the progenitor SMBHs were in a precessing configuration. We use \gwModel{} as our recoil-kick prescription. In all cases, the priors are shown in gray.}
    \label{fig:RSMBH_vs_3C186}
\end{figure}

\section*{Supplemental material}
\label{sec:appendix}

\begin{figure}
    \centering
    \includegraphics[width=\columnwidth]{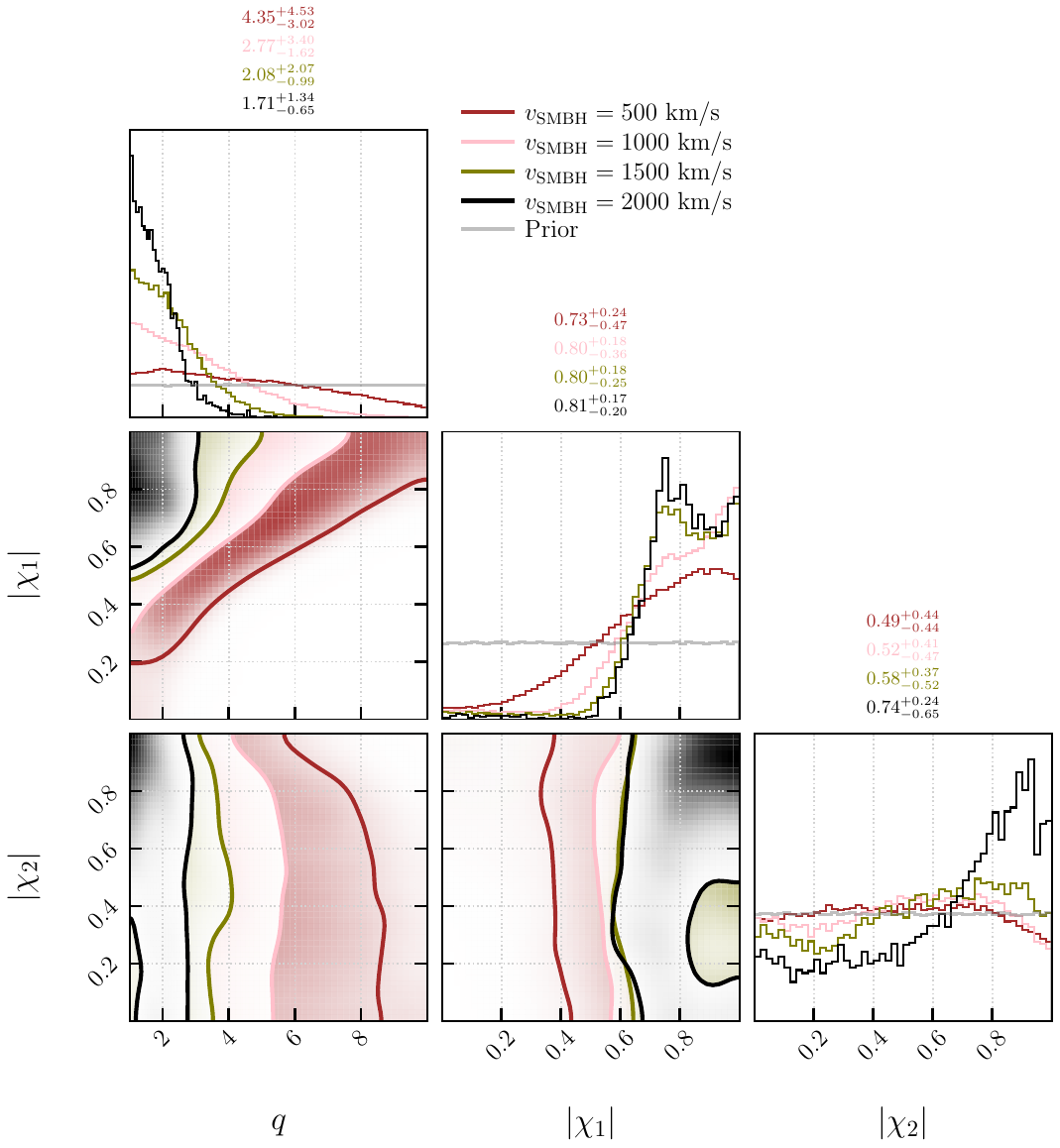}
    \includegraphics[width=\columnwidth]{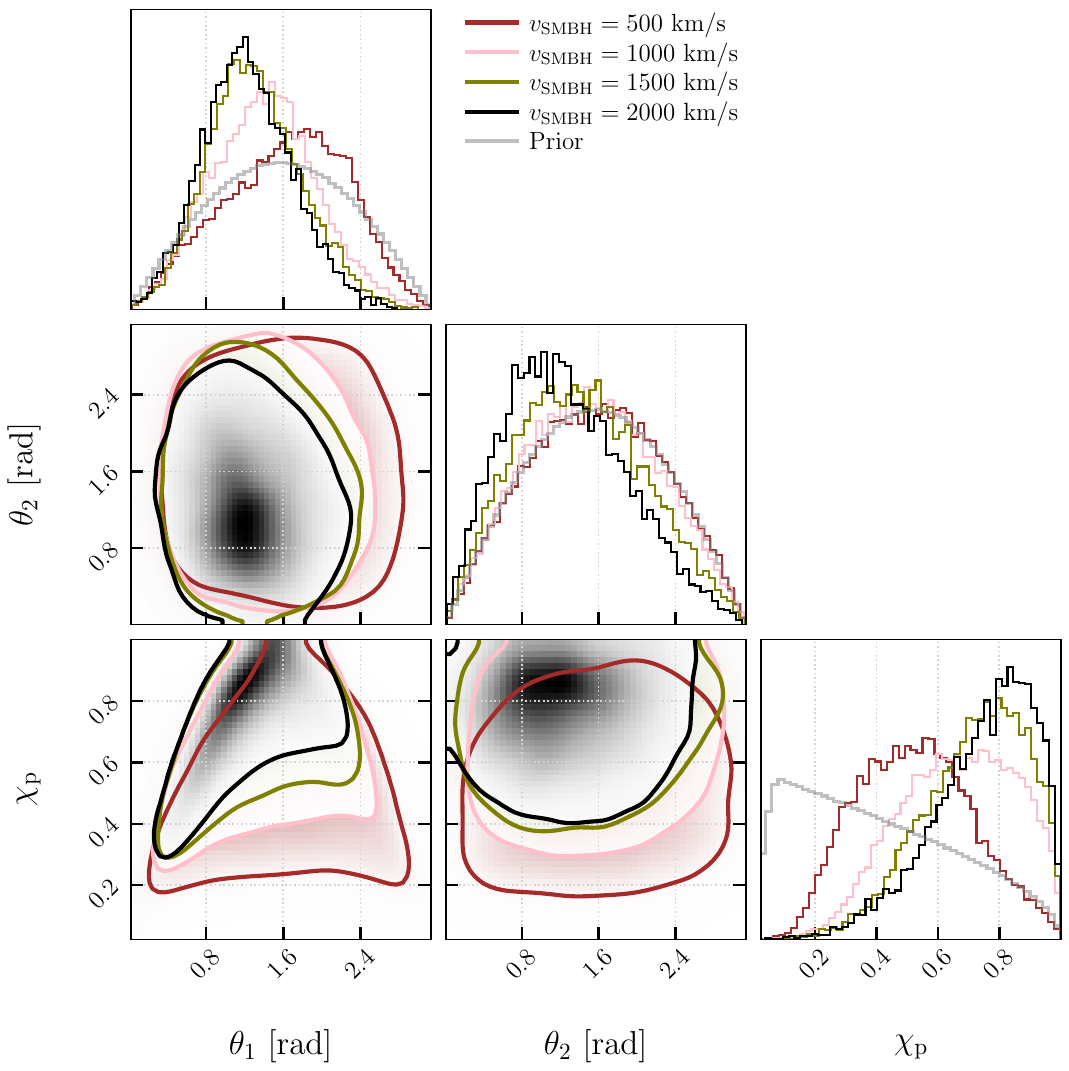}
    \caption{\textbf{Top: }We show the progenitor SMBH mass ratio $q$ and the dimensionless spin magnitudes of the progenitor SMBHs $|\chi_{1,2}|$ consistent with runaway velocities of $[500, 1000, 1500, 2000],\mathrm{km,s^{-1}}$ under the assumption that the progenitor SMBHs were in a precessing configuration. We use \gwModel{} as our recoil-kick prescription. \textbf{Bottom: }We show the angles between the progenitor SMBH spins and the orbital angular momentum of the binary, $\theta_1\ (=\cos^{-1}[\hat{L}\cdot\hat{S}_1])$ and $\theta_2$, together with the effective precession parameter $\chi_{\rm p}$, consistent with those runaway velocities. Note that \gwModel{} does not provide these quantities because it marginalizes over spin angles. We therefore use the \NRSur{} model. In all cases, the priors are shown in gray.}
    \label{fig:different_runaway_speed}
\end{figure}

\noindent \textbf{\textit{Choice of the reference frame: }}For the spin magnitudes $|\chi_{1,2}|$ and angles $(\theta_{1},\theta_{2},\phi_{1},\phi_{2})$, we use a reference frame close to merger. For \NRSur{}, this corresponds to $t = -100M$ before merger, where $M$ is the total mass of the binary, while for \HLZ{} it corresponds to a separation of $\sim 10M$. Note that an isotropic or uniform spin-angle distribution remains isotropic or uniform throughout the binary evolution up to merger~\cite{Gerosa:2015tea}; therefore, these small differences in reference frames do not affect our interpretation.

\noindent \textbf{\textit{Comparison with quasar 3C~186: }}Another recent runaway-SMBH candidate is the quasar 3C~186~\cite{2017A&A600A57C,Chiaberge:2025ouh} at redshift $z=1.068$, nearly the same redshift as RBH-1, for which recent spectroscopic observations have provided strong support for the GW recoil interpretation~\cite{Chiaberge:2025ouh}. Hosted by a massive $\sim 10^{11},M_\odot$ galaxy, 3C~186 has a precisely measured runaway velocity of $1310^{+21}_{-21},\mathrm{km,s^{-1}}$, inferred from Subaru/SWIMS and VLT/XSHOOTER observations by comparing the spatial offset ($11$--$12$ kpc) between the quasar point source and the center of its host galaxy. This runaway velocity is $\sim 200,\mathrm{km,s^{-1}}$ larger than the inferred velocity of RBH-1. Alternative explanations for 3C~186 include a compact binary SMBH, asymmetric broad-line region (BLR) kinematics, extreme nuclear outflows, or an old merger remnant~\cite{Marziani:2025kzd,2022ApJ931165M}. Nonetheless, the GW recoil scenario remains the most probable interpretation~\cite{Chiaberge:2025ouh,Boschini:2024tls,Lousto:2017uav}, though it is not yet definitive.

We follow the same procedure described earlier to identify the region of parameter space capable of producing such a high recoil velocity, under the assumption that the quasar is the remnant of a previous SMBH merger (Fig.~\ref{fig:RSMBH_vs_3C186}). We find that approximately $2\%$ of the prior yields a recoil velocity of at least $1289,\mathrm{km,s^{-1}}$ (the lower limit for the 3C~186 runaway velocity) when \gwModel{} is used. This fraction increases to $\sim 4.4\%$ and $\sim 6.2\%$ when using \HLZ{} and \NRSur{}, respectively.

Next, we obtain the posterior under the assumption of a precessing BBH merger. We restrict ourselves to \gwModel{}, since the other prescriptions yield qualitatively similar conclusions. Because of the extremely large recoil velocity, similar to RBH-1, the progenitor of 3C~186 must have been a precessing SMBH binary with a mass ratio in the range $q \in [1,4]$. However, since the recoil velocity of 3C~186 is $\sim 200,\mathrm{km,s^{-1}}$ larger, the progenitor mass ratio is expected to be even closer to unity. As shown in Fig.~\ref{fig:RSMBH_vs_3C186}, this expectation is borne out: 3C~186 favors a mass ratio of $q = 2.28^{+2.48}_{-1.16}$, whereas RBH-1 favors $q = 2.91^{+3.61}_{-1.75}$.

For the dimensionless spin magnitude of the larger progenitor SMBH, we find that both candidates favor broadly similar values, with 3C~186 exhibiting a mild bimodality at high spins. Overall, the two runaway-SMBH candidates require comparable spin configurations. These results qualitatively match the findings presented in Refs.~\cite{Chiaberge:2025ouh,Boschini:2024tls,Lousto:2017uav}.

\noindent \textbf{\textit{Effect of runaway speed on the SMBH progenitor parameters:}}
Understanding how the inferred progenitor parameters depend on the runaway velocity is important for rapidly characterizing future recoiling SMBH candidates. To illustrate this dependence, we compute progenitor parameters for four representative recoil velocities of $[500, 1000, 1500, 2000],\mathrm{km,s^{-1}}$, assuming a conservative $5\%$ uncertainty on each measurement. We use the same priors as before. For each case, we evaluate the posteriors of the key parameters $q$, $\chi_1$, and $\chi_2$ using \gwModel{} (top panel of Fig.~\ref{fig:different_runaway_speed}). We find that higher recoil velocities push the mass ratio toward the equal-mass limit, while the posterior on $\chi_1$ narrows significantly. Once the recoil velocity exceeds $1500,\mathrm{km,s^{-1}}$, the posterior on $\chi_2$ departs from the prior and shifts toward higher spins.

To assess the impact on spin-precession parameters, we repeat this analysis using \NRSur{}. With increasing recoil velocity, the effective precession parameter $\chi_{\rm p}$ progressively departs from the prior and approaches its maximal value of unity (bottom panel of Fig.~\ref{fig:different_runaway_speed}). Above $1500,\mathrm{km,s^{-1}}$, the posterior on $\theta_1$ saturates, while $\theta_2$ begins to deviate appreciably from its prior only for recoil velocities $\gtrsim 1500,\mathrm{km,s^{-1}}$.

\begin{figure}
    \centering
    \includegraphics[width=\columnwidth]{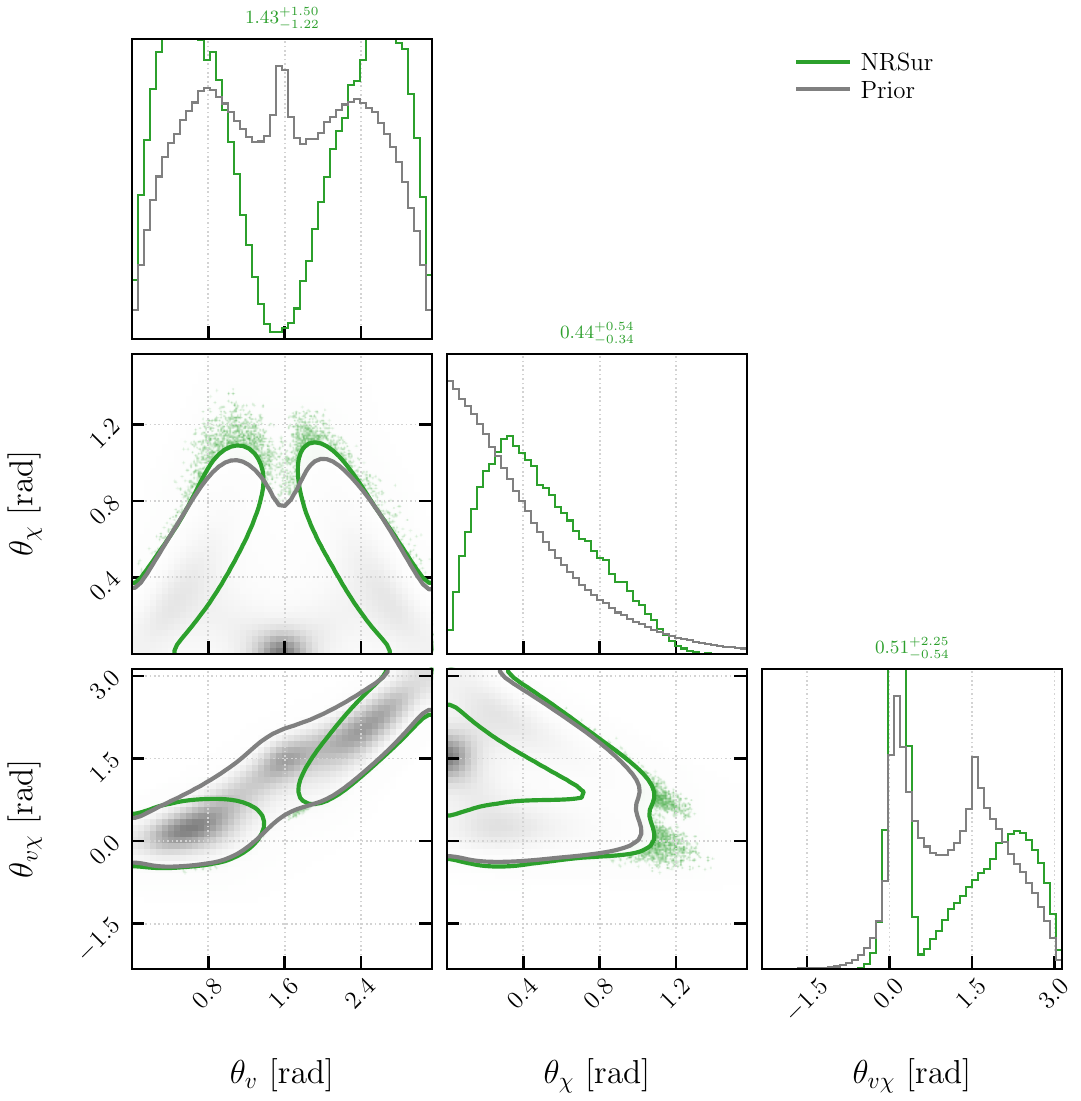}
    \caption{We show the polar angles between the recoil direction and the remnant-spin direction relative to the total angular momentum of the progenitor binary, denoted $\theta_v$ and $\theta_\chi$, respectively, as well as the angle between these two directions, $\theta_{v\chi}$, obtained from \NRSur{}. The priors are shown in gray. If a jet emanating from the RBH-1 were present, its direction would align with the remnant-spin axis. Currently, there is no detection of such a jet in VLA or Chandra data \cite{vanDokkum:2023wed}, but deeper observations may be able to detect a possible faint jet. $\theta_{v\chi}$ represents our prediction of the angle between the recoil direction and the faint jet (which favors an aligned or anti-aligned configuration).}
    \label{fig:kick_angles}
\end{figure}

\begin{figure}
    \centering
    \includegraphics[width=\columnwidth]{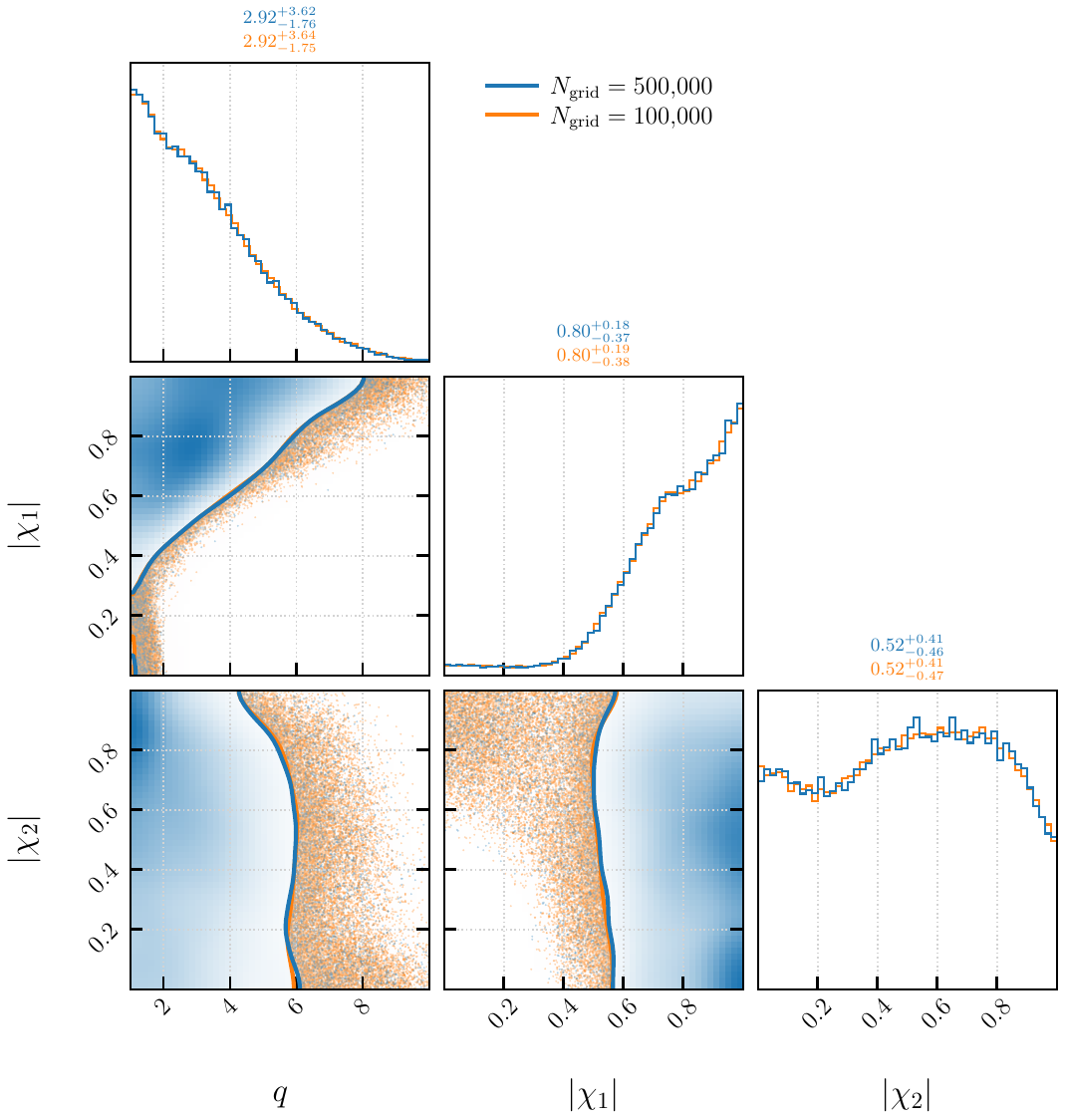}
    \caption{We show the progenitor SMBH mass ratio $q (=m_1/m_2)$ and the dimensionless spin magnitudes $|\chi_{1,2}|$ that are consistent with the inferred RBH-1 speed of $954^{+110}_{-126}\,\mathrm{km\,s^{-1}}$. We compare results obtained using two different numbers of Monte Carlo samples: $100{,}000$ (orange) and $500{,}000$ (blue). We use \gwModel{} as our recoil-kick prescription. The inferred distributions remain practically unchanged as the number of samples is increased, indicating convergence of the analysis.}
    \label{fig:convergence}
\end{figure}

\begin{figure}
    \centering
    \includegraphics[width=\columnwidth]{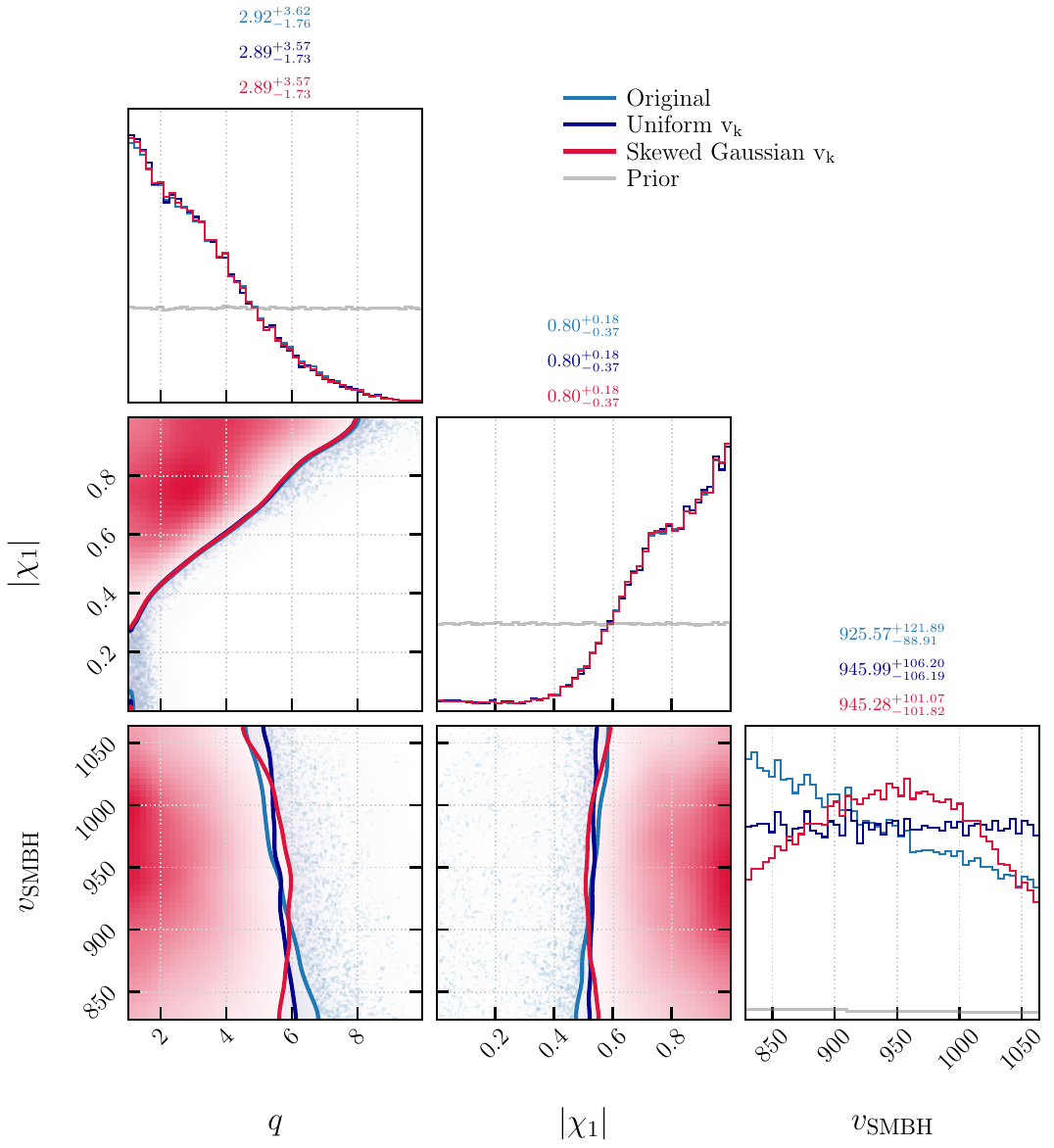}
    \caption{We show the progenitor SMBH mass ratio $q$, the dimensionless spin magnitude of the larger progenitor SMBH $|\chi_{1}|$, and the resulting runaway speed $v_{\rm SMBH}$ consistent with the inferred speeds of $954^{+110}_{-126}\,\mathrm{km\,s^{-1}}$ under three runaway speed distribution for RBH-1: original (blue) and uniform (dark blue), and skewed Gaussian (crimson). We use \gwModel{} as our recoil-kick prescription. In all cases, the priors are shown in gray.}
    \label{fig:comparison_vk_weights}
\end{figure}

\noindent \textbf{\textit{Angle between SMBH velocity and spin: }}If the progenitor black holes merged in a gas-rich environment, a jet could have formed prior to merger. Such a jet would likely align with the spin direction of the remnant SMBH. We therefore estimate the polar angles of the recoil and remnant-spin directions, $\theta_v$ and $\theta_\chi$, relative to the initial total angular momentum of the binary, as well as their mutual angle $\theta_{v\chi}$, corresponding to the angle between a possible jet and the recoil direction. Using \NRSur{}, we find that the runaway velocity of RBH-1 alone is insufficient to constrain these recoil angles (and therefore any possible jet direction) with meaningful precision (Fig.~\ref{fig:kick_angles}), as the posteriors do not differ significantly from the priors. A similar analysis cannot yet be performed with \gwModel{} or \HLZ{}, since they do not currently model recoil directions. Work is underway to incorporate this capability into \gwModel{}, and results will be reported in future work.

\noindent \textbf{\textit{Robustness of analysis: }}A potential concern in Monte Carlo forward modeling with rejection sampling is whether the inferred distributions are converged. To test this, we repeat the analysis using \gwModel{} with $100{,}000$ and $500{,}000$ Monte Carlo samples. We find no appreciable difference in the inferred progenitor properties consistent with the observed recoil velocity of $954^{+110}_{-126},\mathrm{km,s^{-1}}$ (Fig.~\ref{fig:convergence}), demonstrating convergence of our results. We also verify convergence with importance sampling, weighting each prior sample without discarding any, and find unchanged results.

No posterior distribution for the RBH-1 recoil velocity has yet been published~\cite{vanDokkum:2023wed,vanDokkum:2025bah}. We therefore do not assign statistical weights within the allowed velocity range. Instead, we identify the region of SMBH binary parameter space capable of producing recoil velocities consistent with the inferred value of $954^{+110}_{-126},\mathrm{km,s^{-1}}$. For completeness, we also test two reasonable weighting choices: equal weights across the allowed interval and a skewed Gaussian peaked at $954,\mathrm{km,s^{-1}}$, motivated by the asymmetric uncertainty. Reweighting the inferred posteriors under either assumption produces almost no difference from our default results, even though the velocity distribution within the allowed range changes slightly (Fig.~\ref{fig:comparison_vk_weights}).

\noindent \textbf{\textit{Details of the LISA SNR estimate: }}We estimate the characteristic LISA SNR of the progenitor SMBH merger using the \texttt{IMRPhenomXPHM}~\cite{Pratten:2020ceb} waveform model generated with \texttt{lalsimulation}~\cite{lalsuite}. We include the $(2,2)$, $(2,1)$, $(3,3)$, $(3,2)$, and $(4,4)$ spherical harmonic modes together with their negative-$m$ counterparts. The analytic LISA sensitivity curve is taken from Ref.~\cite{Robson:2018ifk}. For a given inclination angle $\iota$, we compute the polarization-combined SNR as
\begin{equation}
\rho^2 = 4 \int \frac{|h_+(f)|^2 + |h_\times(f)|^2}{S_n(f)}, df ,,
\end{equation}
where $h_+(f)$ and $h_\times(f)$ are the frequency-domain waveform polarizations and $S_n(f)$ is the LISA noise power spectral density.

We generate $100$ isotropically distributed inclinations by sampling $\cos\iota$ uniformly in the interval $[-1,1]$. Averaging over these inclinations gives $\left\langle \rho^2 \right\rangle_\iota^{1/2} \simeq 2.8\times10^3$.
Across the sampled inclinations, we find SNRs ranging from $\sim 2.0\times10^3$ to $\sim 3.9\times10^3$. Note that this estimate does not include the full time-dependent LISA antenna response or sky-position dependence, and should therefore be interpreted as a characteristic SNR estimate.

\end{document}